\documentclass[%
 ,twocolumn%
 ,secnumarabic%
,amssymb, amsmath,nobibnotes, prl, aps]{revtex4}
\usepackage{times}%NOTE: Include for submission to arXiv. times fonts NOT available on bug!

\newcommand{\bk}{{\bf k}}

\newcommand{\br}{{\bf r}}

\newcommand{\psib}{{\bar{\psi}}}

\newcommand{\half}{\frac{1}{2}}

\begin{document}

\title{Charge modulation, spin response, and 
%vortex-antivortex fluctuations
dual Hofstadter butterfly
in high-T$_c$ cuprates}%
\author{Zlatko Te\v sanovi\'c}
\address{Department of Physics and Astronomy, Johns Hopkins
University, Baltimore, MD 21218, USA 
\\ {\rm(\today)}
}
\begin{abstract}
\medskip
The modulated density of states observed
in recent STM experiments in underdoped cuprates is argued to be
a manifestation of the charge 
density wave of Cooper pairs (CPCDW). CPCDW formation
is due to superconducting phase fluctuations
enhanced by Mott-Hubbard correlations near half-filling. 
The physics behind the CPCDW is related to a Hofstadter problem in
a dual superconductor. 
It is shown that CPCDW does not impact nodal
fermions at the leading order.
An experiment is proposed
to probe coupling of the CPCDW to 
the spin carried by nodal quasiparticles.

\end{abstract}

\maketitle
Recent 
%scanning tunneling microscopy (STM) 
STM experiments \cite{yazdani,davis,kapitulnik} 
have reinvigorated the debate \cite{zhang,dhlee} on 
the nature of the pseudogap state in underdoped cuprates. 
The central issue is whether the pseudogap state is a phase
disordered superconductor \cite{emerykivelson,balents,qed,herbut}
or some other, entirely different form
of competing order, originating from the
particle-hole channel \cite{dhlee,ddw,sachdev,stripes}.
The observed modulation in the local density of
states (DOS), which
breaks the lattice translational symmetry of CuO$_2$ planes, 
is conceivably attributable to both.

Within the phase-fluctuating superconductor scenario
a natural temptation is to ascribe this modulation to
the ``helium physics'': a system of bosons (Cooper pairs) 
with short range repulsion is superfluid
in its ground state as long as 
it is {\em compressible} \cite{reatto} -- the
only alternative to superfluidity is 
an {\em incompressible} state \cite{fisher}. 
In cuprates, as doping $x$ is reduced toward
half-filling, $x\to 0$, strong onsite repulsion
suppresses particle density fluctuations and reduces compressibility. 
This leads to enhanced phase
fluctuations and reduced superfluid density $\rho_s$, courtesy of the
uncertainty relation $\Delta N\Delta\varphi\agt 1$. 
At $x=x_c$, a compressible superfluid turns
into an incompressible Mott insulator. Such insulator
tends to maintain a fixed number of particles in a given
area and, at some doping $x<x_c$, 
the CuO$_2$ lattice symmetry typically will be 
broken in favor of a superlattice
with a large unit cell, tied to $1/x \gg 1$.
In this Letter I succumb to this temptation and 
examine several of its experimental consequences. 

The first step is to recognize that the pseudogap physics 
{\em differs} in an essential way from the above $^4$He analogy:
cuprates are $d$-wave superconductors and, in contrast to $^4$He
or $s$-wave systems, any useful description must 
contain not only the bosons (Cooper pairs) but also 
{\em fermionic} excitations in the form of nodal 
Bogoliubov-deGennes (BdG) quasiparticles. The quasiparticles 
carry well-defined {\em spin} $S=\frac{1}{2}$ and their
coupling to the charge sector, dominated
by the $S=0$ Cooper pairs, is arguably the crucial element
of quantum dynamics of cuprates. This spin-charge interaction
is topological in origin and peculiar for fluctuating
spin-singlet superconductors \cite{qed}.

The nodal fermions convey another fundametal
information: Cooper pairs in cuprates are inherently
the {\em momentum-space} objects in contrast to the
{\em real-space} pairs behaving as ``elementary'' bosons, like $^4$He or the 
SO(5) hard-core plaquette bosons \cite{zhang}. Thus 
one encounters in cuprates an
echo of the historical debate on Blatt-Schafroth versus BCS pairs.
This is an important issue -- while certain long-distance
features of the two descriptions are equivalent, 
many crucial physical properties are not. 
In particular, the observed charged
modulation is a finite wavevector, non-universal phenomenon.
As shown in this Letter, the modulation patterns and stable states
arising from the two descriptions are essentially different.

To appreciate this difference, note that
Cooper pairs in nodal $d$-wave superconductors are highly non-local
objects in real space and any description in terms of their center-of-mass
coordinates will reflect this non-locality through
complicated intrinsically multi-body, extended-range interactions.
Such complexity haunts any attempt at
constructing a theory using Cooper pairs as
``elementary'' bosons.
The basic idea advanced in this Letter is that, under these circumstances,
the role of ``elementary'' bosons should be accorded to {\em vortices}
instead of Cooper pairs. Vortices in cuprates, with their small cores,
are simple real space objects and the effective theory of quantum fluctuating
vortex-antivortex pairs can be written in the form that is 
local and simple to analyze. 

I start by proposing that the modulation observed in
\cite{yazdani,davis,kapitulnik} reflects 
the Copper pair charge-density wave (CPCDW) in
a fluctuating nodal $d$-wave superconductor.
I then show that the physics behind
CPCDW relates to an Abrikosov-Hofstadter problem \cite{hofstadter,abrikosov} 
for a {\em dual} type-II lattice superconductor with a 
flux per unit cell $f=(1-x)/2$.
This mapping allows one to identify stable 
states as function of $x$ and to 
extract the periodicity and orientation
of CPCDW relative to the CuO$_2$ lattice. 
I elucidate the origin of stable fractions
and contrast the results with those for
the real-space pairs. The
two differ in a fundamental way, akin to the difference between
strongly type-II and strongly type-I superconductors.
Next, I argue that the formation of CPCDW 
is {\em irrelevant} for the physics
of nodal fermions -- CPCDW is a ``high-energy'' phenomenon
in the parlance of the effective theory \cite{qed,herbut}. 
Consequently, the leading behavior of nodal fermions remains
undisturbed. Finally, I suggest an
observable effect of CPCDW which
probes an essential element of the theory:
the presence of a gauge field which 
frustrates the propagation of spin, exclusively carried
by nodal quasiparticles. The effect is an 
{\em enhanced modulation}, with the 
periodicity related to CPCDW, of
the subleading, $T^2$, term in the spin 
susceptibility $\chi$. This effect takes place in the ``supersolid'' 
state, where superconductivity
and CPCDW coexist, and
its experimental observation 
would provide direct evidence of the topological coupling between
the fluctuating vortex-antivortex pairs responsible for
CPCDW and the electronic {\em spin}.

The effective theory of a quantum
fluctuating $d_{x^2-y^2}$-wave superconductor was derived in
\cite{qed} and represents the interactions of 
fermions with $hc/2e$ vortex-antivortex
excitations in terms of two gauge fields,
$v_\mu$ and $a_\mu$:
\begin{equation}
{\cal L}= \bar\Psi[D_0 +iv_0 +
\frac{({\bf D} +i{\bf v})^2}{2m}-\mu]\Psi -
i\Delta\Psi^T\sigma_2 \hat\eta\Psi + {\rm c.c.} 
+ {\cal L}_0~,
\label{lagrangian} 
\end{equation}  
where $\bar\Psi = (\bar\psi_\uparrow,\bar\psi_\downarrow)$, 
$\mu=(\tau,x,y)$,
$D_\mu =\partial_\mu +ia_\mu\sigma_3$,  $\sigma_i$'s
are the Pauli matrices,
and $\hat\eta\equiv D_x^2 - D_y^2$.
${\cal L}_0 [v,a]$ is the
Jacobian of the transformation from discrete (anti)vortex coordinates
to continuous fields $v$ and $a$:
\begin{equation}
\int {\cal D}[\Phi,\phi_s,A_d,\kappa]{\cal C}^{-1}
{\rm e}^{\{\int d^3x(i2A_d\cdot (\partial\times v)+
i2\kappa\cdot (\partial\times a)-{\cal L}_d)\}}~~,
\label{jacobian}
\end{equation}
where ${\cal L}_d[\Phi,A_d,\kappa]$ is
a dual Lagrangian:
\begin{equation}
{\cal L}_d =m_d^2|\Phi |^2 + |(\partial 
- i2\pi A_{d})\Phi |^2 + \frac{g}{2}|\Phi|^4
+ |\Phi|^2(\partial\phi_s-2\pi\kappa)^2,
\label{duallagrangiani}
\end{equation}
and ${\cal C}[|\Phi |]$ is a normalization factor
\begin{equation}
{\cal C}=\int{\cal D}[a,\phi_s,\kappa]
{\rm e}^{\{\int d^3x(i2\kappa\cdot(\partial\times a)
+|\Phi|^2(\partial\phi_s-2\pi\kappa)^2)\}}~.
\label{normalization}
\end{equation}

%Despite its menacing appearance, 
The physics
behind (\ref{lagrangian}) is simple: The fermionic part of ${\cal L}$ is
just the BdG action for a nodal
$d$-wave superconductor, the awkward
phase factor $\exp(i\varphi(x))$ having been removed from
$\Delta$ by a gauge transformation. This transformation 
generates gauge fields
$v$ and $a$, which mimic the effect of vortex-antivortex pair 
fluctuations on the BdG quasiparticles -- $v$ in the charge and $a$ in
the spin channel.  Finally, a bosonic field $\Phi$ 
describes quantum vortex-antivortex pairs:
vortices/antivortices can be thought of as particles/antiparticles
created and annihilated by dual field $\Phi$. The dual ``normal''
state ($\langle\Phi\rangle=0$)
is a physical superconductor while dual
condensate ($\langle\Phi\rangle\not =0$) describes the pseudogap state. 
The purpose behind the mathematics is to reformulate the problem 
in terms of the BdG action for fermions
(\ref{lagrangian}) and the local Lagrangian of vortex bosons 
${\cal L}_d$ (\ref{duallagrangiani}), kept in mutual
communication via two pairs of gauge fields $(v,a)$ and $(A_d,\kappa)$.

Why is this reformulation useful? 
The CPCDW, an intractably non-local problem
in the language of electrons, has a simple local
expression in the dual language of vortex field $\Phi$.
To recognize this, observe that the phase $\varphi (x)$ 
couples in ${\cal L}$ (\ref{lagrangian},\ref{jacobian}) to the overall
electron density as $\frac{i}{2}\bar n\partial_\tau\varphi$, 
where $\bar n=\bar n_\uparrow + \bar n_\downarrow$.
This translates to a dual ``magnetic field'' 
$\nabla\times {\bf A}_d =\half\bar n$ \cite{fisherlee}
seen by vortex-antivortex pairs \cite{footi}. 
Physically,  this effect gives dual voice to the ``helium physics''
discussed earlier: to prevent superfluid ground state the system
turns into an incompressible solid, a dual Abrikosov
lattice \cite{abrikosov,balents}. Therefore, the quantum vortex-antivortex
unbinding leads to the breaking of (lattice) translational
symmetry. When the pattern of symmetry breaking is determined
by the local dual problem, the results are
``communicated'' back to the fermionic part of ${\cal L}$ (\ref{lagrangian})
via the gauge fields $(v,a)$ and $(A_d,\kappa)$ -- hence CPCDW.

The above arguments are explicit in the dual mean-field
approximation, combined with the linearization of
the spectrum near the nodes. The linearization
splits the fermions in (\ref{lagrangian}) into low-energy
nodal spin-$\frac{1}{2}$ Dirac-like
particles $\psi_{\sigma,\alpha}$, where $\alpha=$ 1, $\bar 1$, 2 and $\bar 2$,
and high-energy anti-nodal fermions combined into spin-singlet
Cooper pairs, $\psi_{\sigma,\langle\alpha\beta\rangle}$,
where $\langle\alpha\beta\rangle =$ $\langle 12\rangle$, $\langle 2\bar1\rangle$, $\langle \bar 1\bar 2\rangle$, and $\langle \bar 2 1\rangle$. 
Nodal Dirac fermions have no overall charge density -- the
overall charge is carried 
by $\psi_{\sigma,\langle\alpha\beta\rangle}$ (Cooper pairs). 
Furthermore, $\psi_{\sigma,\langle\alpha\beta\rangle}$ form
spin-singlets and do not couple to $a$.
This enables us to separate the mean-field equations for the spin sector
from those for charge:  
\begin{eqnarray}
\pi\langle n_\uparrow (\br,\tau) +n_\downarrow (\br,\tau)\rangle
=\nabla\times{\bf A}_d(\br)~~,\nonumber\\
\pm\partial_{y(x)}\delta v_0(\br) = \pi {\bf j}^\Phi_{x(y)}(\br)~~,\nonumber \\
m_d^2\Phi - (\nabla - i{\bf A}_d)^2\Phi + g|\Phi|^2\Phi =0~~,\nonumber \\
\langle\frac{\delta {\cal L}}{\delta\Delta (\br)}\rangle
 = (2/\lambda_{\rm eff})\Delta (\br)~~,
\label{meanfield}
\end{eqnarray}
where $n_{\sigma}(x)=\bar\psi_{\sigma,\langle\alpha\beta\rangle} (x)\psi_{\sigma,\langle\alpha\beta\rangle}(x) + \bar\psi_{\sigma,\alpha}\psi_{\sigma,\alpha}$, 
$j_\mu^\Phi$ is a dual current,
$j_\mu^\Phi = -i\Phi^*\partial_\mu\Phi + {\rm c.c.} + A_{d\mu}|\Phi|^2$,
and $\lambda_{\rm eff}$ is the effective coupling constant 
(the last equation is the BdG self-consistency
condition for the pseudogap).

The first of Eqs. (\ref{meanfield}) is an implicit expression for
$\delta v_0(\br)$. In cuprates, the loss of superconductivity through
underdoping is caused by Mott correlations forcing the system
into incompressibility. This suggests that 
the Fourier transform of fermionic compressibility $\chi_c$ at the
reciprocal lattice vector of the charge modulation
is small: $\chi_c({\bf G})\sim x\ll 1$. Thus, 
to a good approximation:
$\langle \delta n(\br)\rangle \approx\chi_c \delta v_0(\br)$. From
the first equation (\ref{meanfield})
$\chi_c \delta v_0(\br) = \frac{1}{\pi}\nabla\times{\bf A}_d(\br) -\bar n$ and
I recast the next two as:
\begin{eqnarray}
\nabla\times (\nabla\times{\bf A}_d(\br)) = \pi^2\chi_c {\bf j}^\Phi (\br)~~,
\label{maxwell}\\
m_d^2\Phi - (\nabla - i{\bf A}_d(\br))^2\Phi + g|\Phi|^2\Phi =0~~. 
\label{gl}
\end{eqnarray}
(\ref{maxwell},\ref{gl}) are the Maxwell and Ginzburg-Landau
equations for a type-II dual superconductor in a dual
field $H_d=\pi\bar n$ ($\kappa_d \sim 1/\sqrt{\chi_c} > 1/\sqrt{2}$, 
since $\chi_c$ is small for low $x$).

The solution of (\ref{maxwell},\ref{gl}) in continuum is just the 
celebrated triangular vortex lattice of Abrikosov \cite{abrikosov}. 
In our dual problem, however, the effect of
the underlying CuO$_2$ lattice must be considered. This is so
since, for dopings of interest, we are quite close to half-filling
and $x=0$. Translated to dual language this means we are
close to $f=1/2$, the maximally frustrated case of (\ref{gl}). 
The pinning to the atomic lattice is significant
and we should expect a near-square symmetry for the resulting CPCDW. 

Eqs. (\ref{maxwell},\ref{gl}) are solved as follows:
\cite{ashot}: first, 
various derivatives in (\ref{maxwell},\ref{gl}) are replaced by
their CuO$_2$ 
lattice counterparts \cite{zlattice}. We then consider (\ref{gl}) with
a uniform dual field $H_d=\half\bar n$ and 
initially set $g\to 0$. 
This is a variant of the Hofstadter problem for dual bosons
$\Phi_i$, with a 
fractional flux $f=p/q=(1-x)/2$ through 
a plaquette of a dual lattice. The solution 
is a ``Hofstadter butterfly'' spectrum
with deep energy minima for major 
fractions \cite{hofstadter}. The ground state is
$q$-fold degenerate and one must choose the linear combination
of states for dual bosons to condense into. The  
degeneracy is lifted by finite $g$ in (\ref{gl}).
Thus, a unique state $\Phi ^{(0)}_i$ is selected, 
the only remaining degeneracy
associated with discrete lattice symmetries.
%transformations relative to the underlying
%dual lattice. 
Once $\Phi ^{(0)}_i$ is known, one computes the
current ${\bf j}^\Phi$ and uses Maxwell equation (\ref{maxwell}) to find
the modulation in dual induction 
$\delta{\bf B}_d={\bf B}_d -{\bf H}_d= \nabla \times{\delta\bf A}_d(\br)$. 
This procedure is then iterated to convergence. 
%In practice,
%as long as $\kappa_d > 1/\sqrt{2}$, the leading
%order solution will suffice.

The major fractions and their modulation patterns 
are {\em primarily} determined by the Abrikosov-Hofstadter problem
(\ref{gl}), the magnetic energy being a small correction
in a type-II system. The interactions
among vortices in $\Phi ^{(0)}_i$ responsible for these patterns
are intrinsically multi-body and of extended-range -- 
they are the interactions among the center-of-mass coordinates of
Cooper pairs. This is in contrast to the
real-space pairs with pairwise short-ranged
interactions $V(\br-\br')$. The pair density-wave is
determined not by (\ref{gl}) but by the
Wigner crystallization, encoded in (\ref{maxwell}), which in this
limit turns to the minimization of: 
$\half\int d^2rd^2r' {\bf B}_d(\br)V(\br-\br'){\bf B}_d(\br')$, where
${\bf B}_d (\br)=\sum_i\delta(\br - \br_i)$ and $\{\br_i\}$ are
the pairs' positions. Thus, the Cooper and 
the real-space pairs correspond to the two {\em opposite} limits of
(\ref{maxwell}, \ref{gl}), that of the type-II
and the type-I regime of a dual superconductor, respectively.
%The density-wave patterns associated with these two limits
%are different and distinguishable by experiments.

While the analysis of (\ref{maxwell},\ref{gl}) is
given in \cite{ashot}, I outline here 
general features of the solution. $\Phi ^{(0)}_i$ and
%the accompanying modulation in
$\nabla \times\delta{\bf A}_d(\br)$ 
break the translational symmetry of the dual and
the CuO$_2$ lattice. The new superlattice is characterized by
the set of reciprocal vectors $\{{\bf G}_i\}$. 
The major fractions $f$, i.e.
the energetically most favored states, are those with $q$ being
a small integer, (integer)$^2$ or a multiple
of 2, reflecting the square symmetry of the CuO$_2$ planes.
In the window of $x$ relevant to cuprates, these
are $f=7/16$, $4/9$, $3/7$, $6/13$, $11/24$, 
$15/32$, $13/32$, $29/64$, $27/64$, $\dots$, 
($x=0.125~(1/8)$, $0.111~(1/9)$,
$0.143~(1/7)$, $0.077~(1/13)$, $0.083~(1/12)$,
$0.0625~(1/16)$, $0.1875~(3/16)$,
$0.09375~(3/32)$, $0.15625~(5/32)$) etc. 
Other potentially prominent $f$, like $1/4$, $1/3$, $2/5$, or $3/8$, 
correspond to $x$ outside the regime
of vortex-antivortex fluctuations. 

The above information allows insight
into $\{{\bf G}_i\}$'s of major fractions. 
The non-linear term in (\ref{gl}) favors the smallest
unit cell containing an integer number of flux quanta
and a homogeneous modulation in $|\Phi_i|$. 
These conditions single out doping $x=0.125$ ($f=7/16$) 
as a particularly prominent fraction.
At $x=0.125$ ($q=16$), the modulation in $\nabla \times{\bf A}_d(\br)$ can
take advantage of a $4\times 4$ elementary checkerboard block 
which, when oriented along the $x(y)$ direction,  
fits neatly into plaquettes of the dual lattice. 
Near $f=1/2$, however, a large number
of vortices ($p=7$) per such a block leads to a redistribution and 
a larger, rhombic unit cell \cite{ashot} -- the energy gain 
relative to the $4\times 4$ checkerboard,
however, is {\em extremely} small. The modulation in
$\delta\bar n(\br)$ (and $\Delta$) (\ref{meanfield})
still retains a memory of the $4\times 4$ 
block and is characterized by wavevectors
${\bf G_1}=(\pm 2\pi/4a,0)$, ${\bf G_2}=(\pm 2\pi/8a,\pm 2\pi/4a)$, and
${\bf G_3}=(\pm 2\pi/(8a/3),\pm 2\pi/4a)$,
with ${\bf G_1}$ oriented along the {\em antinodal}
(either $x$ or $y)$ directions of the CuO$_2$ lattice. 
The domains of the above modulation pattern 
offer a natural explanation for the observations in Ref. \cite{davis}.

The next leading fractions are $x=0.077~(1/13)$ ($f=6/13$)
and $x=0.111~(1/9)$ ($f=4/9$).
The modulation patterns are now more complicated and do not
fit easily into the underlying CuO$_2$ lattice.
$\delta\bar n(\br)$ (and $\Delta$) \cite{zlattice} exhibits 
a rhombic unit cell with $\{{\bf G}_i\}$'s 
oriented closer to the lattice
diagonals, i.e. the {\em nodal} directions. Thus, as $x$  
decreases away from 1/8 there 
will be a tendency to {\em reorient} the 
superlattice away from antinodal directions and align it
closer to the CuO$_2$ lattice diagonals. Such reorientation effects
of the CPCDW, if observed, would 
provide support for the physics described in this Letter.

The above considerations 
include dopings like $x=1/8$ or $1/9$
for which cuprates are typically still superconducting.
In such cases the mean-field (\ref{meanfield}) is inadequate
and one must include fluctuations in $\Phi$ and $A_d$. The fluctuations
act to depopulate the mean-field ground state $\Phi^{(0)}$ and
transfer some of the dual bosons to the states nearby in 
energy. As $x$ increases toward $x_c$, $\Phi^{(0)}$ eventually 
ceases to be {\em macroscopically} occupied ($\langle\Phi\rangle\to 0$) and
the system returns to the superconducting state. However, as long as
the transition is not strongly first-order, dual bosons still
preferentially occupy the states close to $\Phi^{(0)}$ on the
``Hofstadter butterfly'' energy landscape. This results in
$\langle |\Phi (\br)|^2\rangle$ which is finite and still modulated.
Only for yet higher $x$ will the translational symmetry of the
superconducting state be finally restored.

The above is an example of the 
``supersolid state'',
in which superconductivity coexists with the CPCDW. The 
modulation is dominated by $\Phi^{(0)}$ and thus our mean-field
symmetry analysis of major fractions 
still goes through. The fluctuations that produce the ``supersolid'' state 
consist of a liquid of vacancies and interstitials superimposed
on the original mean-field dual vortex lattice. This leads
to low $\rho_s$ and tends to
shift the periodicity away from 
the mean-field set of $\{{\bf G}_i\}$'s associated with
major fractions, particularly as a function
of $T$, since the self-energies of vacancies and interstitials
are generically different. Such fluctuation-induced incommensurability could
be behind the difference between the CPCDW periodicities
observed in \cite{yazdani} (high $T$) and \cite{davis} (very low $T$). 
%The study of fluctuation 
%corrections to (\ref{meanfield},\ref{gl},\ref{maxwell}) 
%is beyond the scope of this Letter and is 
%best left to a computational analysis \cite{sudbo}.

The preceding discussion of the charge sector sets the stage 
for the question of what happens to {\em spin}, carried
by nodal quasiparticles (for convenience, 
I now rotate a $d_{x^2-y^2}$-wave superconductor into a $d_{xy}$-wave one). 
The CPCDW affects low-energy fermions in two ways:
first, $\delta v_0(\br)$ couples to $\psi_{\sigma,\alpha}$ as a periodically
modulated chemical potential and can be
absorbed into a locally varying Fermi wavevector:
$k_F\to k_F +\delta k_F (\br)$, where $\delta k_F (\br) =\delta v_0(\br)/v_F$.
Such shift leaves the nodal point in the energy space undisturbed.
Similarly, there also is a modulation in the size of the
pseudogap: $\Delta\to\Delta + \delta\Delta (\br)$, arising from the
BdG self-consistency equation (\ref{meanfield}). 
Near the nodes $\Delta ({\bf P};\bk)
\to \Delta (\hat k^2_x-\hat k^2_y)+\delta\Delta (\{{\bf G}_i\},{\bf k})$,
where ${\bf P}$ is related to
the center-of-mass momentum of Cooper pairs. 
Assuming that the pseudogap retains the overall $d_{x^2-y^2}$-wave symmetry
throughout the underdoped regime, one finds 
$\delta\Delta (\{{\bf G}_i\},{\bf k})\sim \hat k^2_x-\hat k^2_y $.
Again, the nodal point is left
intact.  The semiclassical spectrum is:
\begin{equation}
E(\bk;\br) = \pm\sqrt{v_F^2(\br)k_x^2+ v_\Delta ^2(\br)k_y^2}~~,
\label{spectrum}
\end{equation}
where $v_F(\br)= v_F + (\delta k_F (\br)/m)$ and
$v_\Delta (\br)=v_\Delta + (\delta\Delta (\br)/k_F)$ \cite{footnote}. 
The local DOS exhibits modulation at wavevectors $\{{\bf G}_i\}$'s
but still vanishes linearly at the nodes.
The only exception to this behavior is the situation in which
CPCDW is {\em commensurate} with the nodes
and $\{{\bf G}_i\}$'s happen to coincide with some
of the internodal wave vectors:
${\bf Q}_{1\bar 1}$, ${\bf Q}_{1\bar 2}$, etc. 
Such commensuration can only be 
purely accidental since the dual lattice physics (\ref{maxwell},\ref{gl})
that determines $\{{\bf G}_i\}$'s has no simple relation to the location
of nodes in the Brillouin zone.

There is, however, a yet another way by which the CPCDW 
affects nodal fermions $\psi_{\sigma,\alpha}$. 
This is through the coupling to a Berry gauge
field $a_\mu^i$, which describes topological frustration of BdG ``spinons''
moving through space filled with fluctuating $hc/2e$ vortex-antivortex 
pairs. This non-trivial coupling of charge and spin sector is
captured by the effective Lagrangian:
\begin{equation}
{\cal L}_{\rm f}=\psib _n \left(i\gamma_{\mu}\partial_{\mu}+
\gamma_{\mu}a_{\mu} 
\right)\psi_n 
+ {\cal L}_0^a[a_{\mu}]~~, 
\label{qed3}
\end{equation}  
obtained as the low energy ($\ll\Delta$) 
limit of ${\cal L}$ (\ref{lagrangian}).
In (\ref{qed3}), $\psi_{\sigma,\alpha}$ have been 
arranged into four component Dirac-BdG
spinors $\psi_n$ following conventions of Ref. \cite{qed} and the 
summation over $N =2$ nodal flavors is understood.

Below the pseudogap energy scale $\Delta$,
the spin correlator of {\em physical} electrons is \cite{qedthermal}:
\begin{equation}
\langle S_z(-k) S_z(k)\rangle=\frac{\Pi^F_A (k)\Pi^0_A(k)}{\Pi^F_A(k) +\Pi^0_A(k)}
\frac{\bk^2}{\bk^2+\omega_n^2}~~,
\label{chi}
\end{equation} 
where $\Pi^F_A(k)\sim |k|$ denotes the fermion current polarization and
$\Pi^0_A$ is the bare stiffness of $a$ in ${\cal L}_0$. 
In the pseudogap state $a$ is massless and $\Pi^0_A \sim k^2$ dominates
the expression for the static spin susceptibility $\chi$, leading to
a non-Fermi liquid behavior of nodal quasiparticles \cite{qedthermal}.
In the superconducting state, $a$ has 
mass $M^2$, $\Pi^0_A\to M^2$, 
and the leading order behavior is set
by $\Pi^F_A$. For $T \ll \Delta$:
\begin{equation}
\chi \sim
(2N\ln 2/\pi) T - \frac{(2N\ln 2/\pi)^2}{M^2}T^2 +\dots~~. 
\label{chi2}
\end{equation}

The leading term ($\sim T$) in (\ref{chi2}) 
is just the renormalized $d$-wave Yoshida function of non-interacting
BdG quasiparticles. The subleading term
($\sim T^2$), however,  involves $M^2$.
In the ``supersolid'' phase $M^2$ is {\em modulated} 
via the non-uniformity in 
$\langle |\Phi|^2\rangle$ (\ref{duallagrangiani}) --
this modulation carries an imprint of 
the CPCDW periodicity set by $\{{\bf G}_i\}$'s,
since it reflects the variation of $\langle |\Phi|^2\rangle$
on the lattice dual to the CuO$_2$ one. 
Furthermore, since $M^{-2}\sim\xi_{\rm d}$ \cite{qed}, 
where $\xi_{\rm d}$ is the {\em dual} superconducting
correlation length, the 
$T^2$ term in (\ref{chi2}) is $\propto \xi_{\rm d}$
and consequently strongly enhanced as $x\to x_c$. The combination of
modulation and enhancement, as the superconductivity is extinguished
at $x=x_c$, sets this term apart from other contributions to $\chi$.  
The observation of such a modulation, in a $\mu$SR or an NMR
experiment, for instance,
would provide a vivid illustration of the subtle interplay
between the charge and spin channels which is the hallmark
of theory (\ref{lagrangian}).

I thank A. Melikyan, J.C. Davis, M. Franz, J.E. Hoffman, A. Sudb\o , 
O. Vafek, and A. Yazdani for useful discussions.
This work was supported in part by the NSF grant DMR00-94981.


\begin{references}
\bibitem{yazdani} M. Vershinin {\it et al.}, Science {\bf 303}, 1995 (2004). 
\bibitem{davis} K. McElroy {\it et al.}, cond-mat/0404005; J.E. Hoffman
{\it et al.}, Science {\bf 295}, 466 (2002).   
%; Nature {\bf 422}, 592 (2003).
\bibitem{kapitulnik} C. Howald {\it et al.}, \prb {\bf 67}, 014533 (2003).
%\bibitem{corson} J. Corson {\em et al.}, Nature {\bf 398}, 221 (1999).
%\bibitem{ong} Z. A. Xu {\em et al.}, Nature {\bf 406}, 486 (2000);
%Y. Y. Wang {\em et al.}, Science {\bf 299}, 86 (2003).
%\bibitem{campuzano} J. C. Campuzano {\em et al.}, unpublished.
\bibitem{zhang} H.D. Chen, O. Vafek, A. Yazdani, and S.C. Zhang,
cond-mat/0402323; H.D. Chen {\it et al.}, 
%S. Capponi, F. Alet, and S.C. Zhang,
cond-mat/0312660; H. D. Chen {\it et al.}, \prl {\bf 89}, 137004 (2002).
\bibitem{dhlee} H.C. Fu, J.C. Davis, and D.H. Lee, cond-mat/0403001.
\bibitem{emerykivelson} V.J. Emery and S.A. Kivelson, Nature {\bf 
374},
434 (1995).
\bibitem{balents} L. Balents, M.P.A. Fisher and C. Nayak, \prb {\bf 60},
1654 (1999).
\bibitem{qed} M. Franz and Z. Te\v sanovi\' c, \prl {\bf 87}, 257003 
(2001); Z. Te\v sanovi\' c, O. Vafek, and M. Franz, \prb {\bf 65},
180511 (2002);
M. Franz, Z. Te\v sanovi\' c, and O. Vafek, {\it ibid.} {\bf 66}, 054535 
(2002).
\bibitem{herbut} I.F. Herbut, \prl {\bf 88}, 047006 (2002); 
B. Seradjeh and I.F. Herbut, \prb {\bf 66}, 184507 (2002).
\bibitem{ddw} S. Chakravarty, R.B. Laughlin, D.K. Morr, and C. Nayak,
\prb {\bf 63}, 094503 (2001).
\bibitem{sachdev} Y. Zhang, E. Demler, and S. Sachdev, \prb {\bf 66},
094501 (2002).
\bibitem{stripes} J. Zaanen and O. Gunnarson, \prb {\bf 40}, 7391 (1989);
K. Machida, Physica C {\bf 158}, 192 (1989);
S.A. Kivelson {\it et al.}, \rmp {\bf 75}, 1201 (2003).
\bibitem{reatto} L. Reatto and G.V. Chester, Phys. Rev. {\bf 155}, 88 (1967).
%L. Reatto, Phys. Rev. {\bf 183}, 334 (1969).
\bibitem{fisher} M.P.A. Fisher {\it et al.}, \prb {\bf 40}, 546 (1989).
\bibitem{hofstadter} D.R. Hofstadter, \prb {\bf 14}, 2239 (1976). 
\bibitem{abrikosov} A.A. Abrikosov, Sov. Phys. JETP {\bf 5}, 1174 (1957).
%\bibitem{oskar} I thank O. Vafek for illuminating discussions on
%the subject of global SU(2) spin symmetry in QED3 theory \cite{qed}.
\bibitem{fisherlee} This argument is
borrowed from M.P.A. Fisher and D.H. Lee, \prb {\bf 39}, 2756 (1989).
\bibitem{footi} There is no analogue ``magnetic field''
for the spin dual field $\kappa$ since
$\bar n_\uparrow -\bar n_\downarrow=0$.
\bibitem{ashot} A. Melikyan and Z. Te\v sanovi\' c, cond-mat/0408344.
\bibitem{zlattice} A care is needed here since Cooper pairs reside
on bonds. Also, the non-locality in $\chi_c$ complicates the
algebra behind (\ref{maxwell}) while leaving the physics unchanged
\cite{ashot}.
%\bibitem{sudbo} S. Mo, J. Hove, and A. Sudb\o, \prb {\bf 65}, 104501 (2002);
%M. Franz and S. Teitel, \prb {\bf 51}, 6551 (1995).
\bibitem{footnote} The reasoning behind (\ref{spectrum}) 
can be fortified beyond semiclassical:
the modulation $\delta v_0 (\br)$ defines some new ``band-structure'' 
of a lattice $d$-wave superconductor
with an effective hopping $t_{ij}$ and a bond pseudogap
$\Delta_{ij}$. 
If we fold the original Brillouin zone to accommodate 
the supercell of charge modulation, the nodal points are still at the
new Fermi surface, as long as $\Delta_{ij}$ retains the overall $d$-wave
symmetry. Near the nodes, the spectrum is linear with
perturbative velocity renormalizations $\sim (\delta v_0 (\{{\bf G}\}_i)) ^2$.
\bibitem{qedthermal} O. Vafek and Z. Te\v sanovi\' c, 
\prl {\bf 91}, 237001 (2003).

\end{references}
\end{document}